\documentclass[twocolumn,secnumarabic,amssymb, amsmath, nofootinbib,tightenlines,
nobibnotes, aps, prl,epsfig]{revtex4}
\usepackage{graphicx}
\usepackage{dcolumn}
\usepackage{bm}
\begin{document}
\preprint{APS/123-QED}
\title{Longitudinal structure function from logarithmic slopes of $F_{2}$ at low $x$}

\author{G.R.Boroun}%
 \email{grboroun@gmail.com; boroun@razi.ac.ir }
\affiliation{ Physics Department, Razi University, Kermanshah
67149, Iran}
\date{\today}
\begin{abstract}
Using Laplace transform techniques, I calculate the longitudinal
structure function $F_{L}(x,Q^{2})$ from the scaling violations of
the proton structure function $F_{2}(x,Q^{2})$, and make a
critical study of this relationship between the structure
functions at leading order (LO) up to next-to-next-to leading
order (NNLO) analysis at small $x$. Furthermore, I consider heavy
quark contributions to the relation between the structure
functions, which leads to compact formula for
$N_{f}=3+\mathrm{Heavy}$. The nonlinear corrections to the
longitudinal structure function at LO up to NNLO analysis are
shown in the $N_{f}=4$
 (light quark flavor) based on the nonlinear corrections at $R=2$ and $R=4~GeV^{-1}$. The results are compared
 to
 experimental data of the longitudinal proton structure function $F_{L}$ in the
 range of $6.5 \leq Q^{2} \leq 800 ~GeV^{2}$.\\

\end{abstract}
 \pacs{***}
\maketitle
\tableofcontents
\subsection{I. Introduction}
The inclusive deep inelastic scattering (DIS) measurements are of
importance to understanding the gluonic substructure of  proton at
low values of Bjorken variable $x$. The reduced cross section is
defined in the following form
\begin{eqnarray}
\widetilde{\sigma}(x,Q^{2})=F_{2}(x,Q^{2})-\frac{y^{2}}{Y_{+}}F_{L}(x,Q^{2}),
\end{eqnarray}
where $Y_{+}=1+(1-y)^2$, $y={Q^{2}}/{xs}$ is the inelasticity, $s$
is the center-of-mass  squared energy of  incoming electrons and
protons, $F_{2}(x,Q^{2})$ and $F_{L}(x,Q^{2})$ are the
transverse and longitudinal structure functions, respectively.\\
The structure functions describe the momentum distributions of
partons in nucleon. A measurement of the proton structure
functions ($F_{2}$ and $F_{L}$) at low values $x$ is directly
sensitive to the gluon density. This provides a sensitive test for
perturbative QCD (pQCD).\\
The longitudinal structure function is determined by measurements
of differential cross sections at different values of $\sqrt{s}$
at HERA, where data on $\sqrt{s}$ for electron beam energies of
$E_{e}\simeq 27.5 GeV$ and for proton beam energies of $E_{p}=920,
820, 575$ and $460 GeV$ are collected [1-2]. The experimental data
for neutral current were also collected for $0.045\leq Q^{2}\leq
50000 GeV^{2}$ and $6E-7 \leq x \leq 0.65$ at values of the
inelasticity $0.005\leq y \leq0.95$.\\
 The contribution of $F_{L}$
to  reduced cross section ( Eq.(1)) is significant only at high
value of the inelasticity $y$, i.e., the kinematic region. The
latter case corresponds to the low values of the Bjorken variable
$x$, and the longitudinal structure function is related to the
gluon density of the proton. As gluons are the most influencing at
low $x$, therefore the quark contribution to the longitudinal
structure function and singlet structure function is ignored in
next step.\\
The gluonic longitudinal structure function can be written as
\begin{eqnarray}
x^{-1}F_{L}(x,Q^{2}){=}<e^{2}>C_{L,g}(\alpha_{s},x)\otimes
g(x,Q^{2}),
\end{eqnarray}
where $g(x,Q^{2})$ represents the gluon density, and $<e^{2}>$ is
the average squared charge ($=5/18$ for even $N_{f}$, where
$N_{f}$ denotes number of active light flavours). The symbol
$\otimes$ denotes the Mellin convolution according to the usual
prescription.\\
 The perturbative expansion of the gluon
coefficient function can be written as
\begin{eqnarray}
C_{L,g}(\alpha_{s},x)=\sum_{n=1}(\frac{\alpha_{s}}{4\pi})^{n}c_{L,g}^{n}(x),
\end{eqnarray}
where $n$ is the order in running coupling constant.\\
 The reduced
cross section for deep inelastic lepton-nucleon scattering (Eq.1)
is defined  in terms of the proton structure function $F_{2}$. At
low values of $x$, the gluon contribution to the proton structure
function $F_{2}$ dominates over the flavor singlet contribution.
The DGLAP evolution equation for gluon dominating $F_{2}$
structure function is given by
\begin{eqnarray}
\frac{\partial F_{2}^{s}(x,Q^{2})}{\partial
\ln(Q^{2})}=\frac{\alpha_{s}(Q^{2})}{2\pi}P_{qg}(\alpha_{s},x){\otimes}G(x,Q^{2}),
\end{eqnarray}
where $F_{2}^{s}$ is the singlet distribution function. The
splitting function $P_{qg}$ is the leading order (LO) up to
high-loop corrections to the QCD $\beta$-function as
\begin{eqnarray}
P_{qg}(\alpha_{s},x)=P^{LO}_{qg}+\frac{\alpha_{s}(Q^{2})}{2\pi}P^{NLO}_{qg}+...~.
\end{eqnarray}
Several methods to relate $F_{L}$ and $F_{2}$ scaling violation to
the gluon density at small $x$ were suggested previously [3-4].
These methods were proposed to isolate the gluon distribution by
its expansion around $z=\frac{1}{2}$. In LO analysis with
$N_{f}=4$, the authors of Ref.[3] suggested an approximate
relation between the gluon density at the point $2.5x$ and the
longitudinal structure function $F_{L}$ at the point $x$ in the
following form
\begin{eqnarray}
F_{L}(x,Q^{2})=\frac{2\alpha_{s}}{\pi}\frac{\sum_{i=1}^{N_{f}}e_{i}^{2}}{5.9}G(2.5x,Q^{2}).
\end{eqnarray}
Equation (6) was derived in an expansion of gluon distribution
around $z=\frac{1}{2}$. A similar relation for derivation of
$F_{2}(x,Q^{2})$ with respect to ${\ln}Q^{2}$ based on the
expansion of the gluon distribution around $z=\frac{1}{2}$ was
found in [4], where the following result was obtained
\begin{eqnarray}
\frac{\partial F_{2}(x,Q^{2})}{\partial \ln Q^{2}
}=\frac{5\alpha_{s}}{9\pi}\frac{2}{3}G(2x,Q^{2}).
\end{eqnarray}
Combining Eqs.(6) and (7), one could calculate the longitudinal
structure function by derivation of the structure function at a
rescaled value $\frac{\zeta_{2}}{\zeta_{L}}x$, where
$\zeta_{2}\simeq 0.5$ and $\zeta_{L}\simeq 0.4$. The corresponding
LO expression is
\begin{eqnarray}
F_{L}(x,Q^{2})=\frac{\partial F_{2}(\eta x,Q^{2})}{\partial \ln
Q^{2} },
\end{eqnarray}
where $\eta \simeq 1.25$.\\
In addition, two different methods  were suggested [5-6],  the
 derivatives of the
structure functions were based in the expansion of the gluon
distribution around the arbitrary point $z=\alpha$. The results
were derived at an arbitrary point of expansion as follows
\begin{eqnarray}
F_{L}(x,Q^{2})=\frac{10\alpha_{s}}{27\pi}G(\frac{x}{1-\alpha}(\frac{3}{2}-\alpha),Q^{2}),
\end{eqnarray}
and
\begin{eqnarray}
\frac{\partial F_{2}(x,Q^{2})}{\partial \ln Q^{2}
}=\frac{10\alpha_{s}}{27\pi}G(\frac{x}{1-\alpha}(\frac{3}{2}-\alpha),Q^{2}).
\end{eqnarray}
Eqs. (9) and (10)  strongly depend on the momentum fraction
carried by gluons in Eqs. (2) and (4), and show the behavior of
the gluonic
 structure functions based on the expansion of the
gluon distribution around $z=\alpha$.\\
 In this paper I introduce a method to calculate structure functions using the Laplace
transform techniques.  The paper is organized as follows. In
section II, I find the relation between the structure functions at
small $x$ at LO analysis. In section III, I consider the
high-order corrections to the relation between the structure
functions. In section IV, I utilize obtained solution to calculate
the nonlinear behavior of the longitudinal structure function at
hot-spot point at LO analysis, and present an analytical analysis
of the longitudinal structure function. Then I compare the
obtained result with H1 experimental data. In section V, I study
the high-order corrections to the nonlinear behavior of the
longitudinal structure function. My conclusion is given in section
VI. In Appendix A, I present the results for the splitting
functions and coefficients in the inverse- Laplace transform
method at some values of $Q^{2}$. Appendix B includes the
analytical expression for $F^{\gamma p}_{ 2}(x,Q^{2})$. In
Appendix C and D I present the high-order corrections and
high-order ratios at NLO up to NNLO at small $x$. Appendix E deals
with a technical detail including the inverse- Laplace transform
of the nonlinear
kernels at LO and high order corrections presented in Appendix F.\\


\subsection{II. General Method}
In pQCD, the evolution equations for proton and longitudinal
structure functions are given by in terms of the non singlet,
singlet and gluon coefficient functions. At small values of $x$
the gluon contribution to the structure functions dominate over
the flavour singlet and non-singlet contribution. Therefore, in
this research I consider the gluonic structure functions evolution
equations.\\
 $\textit{A: Four
Flavours:}$ One could write the LO equation for the evolution of
the proton structure function at low values of $x$ as
\begin{eqnarray}
\frac{\partial F_{2}(x,Q^{2})}{\partial \ln(Q^{2})}= \frac{10}{18}
\frac{\alpha_{s}(Q^{2})}{\pi}\int_{x}^{1}\frac{x}{y^{2}}
P_{qg}^{LO}(\frac{x}{y})G(y,Q^{2})dy.
\end{eqnarray}
The longitudinal structure function for gluon dominating is given
by
\begin{eqnarray}
F_{L}(x,Q^{2})=\frac{20}{9}
\frac{\alpha_{s}(Q^{2})}{\pi}\int_{x}^{1}\frac{1}{y}
c^{LO}_{L,g}(\frac{x}{y})G(y,Q^{2})dy.
\end{eqnarray}
Considering the coordinate transformation as $\nu=\ln(1/x)$ and
$\omega=\ln(1/y)$ [7], one could rewrite Eqs. (11) and (12) with
respect to these variables
\begin{eqnarray}
\mathcal{\widehat{F}}_{2}(\nu,Q^{2})&=&\int_{0}^{\nu}\widehat{G}(\omega)e^{-(\nu-\omega)}(1-2e^{-(\nu-\omega)}\nonumber\\
&&+2e^{-2(\nu-\omega)})d \omega ,
\end{eqnarray}
and
\begin{eqnarray}
\mathcal{\widehat{F}}_{L}(\nu,Q^{2})=\int_{0}^{\nu}\widehat{G}(\omega)e^{-2(\nu-\omega)}(1-e^{-(\nu-\omega)})d
\omega ,
\end{eqnarray}
where $\widehat{f}(\nu,Q^{2})=f(e^{-\nu},Q^{2})$,
$\mathcal{\widehat{F}}_{2}(\nu,Q^{2})=\frac{18\pi}{10\alpha_{s}}\frac{\partial
\widehat{F}_{2}(\nu,Q^{2})}{\partial \ln(Q^{2})}$ and
$\mathcal{\widehat{F}}_{L}(\nu,Q^{2})=\frac{9\pi}{20\alpha_{s}}\widehat{F}_{L}(\nu,Q^{2})$.\\
Defining the Laplace transforms
$\mathcal{\widehat{F}}_{2}(s,Q^{2})={\mathcal{L}}[{\widehat{F}}_{2}(\nu,Q^{2});s]$
and
$\mathcal{\widehat{F}}_{L}(s,Q^{2})={\mathcal{L}}[\mathcal{\widehat{F}}_{L}(\nu,Q^{2});s]$,
explicitly from (13) and (14), one obtains the structure functions
in $s$-space as
\begin{eqnarray}
\mathcal{{F}}_{2}(s,Q^{2})=g(s,Q^{2}){\times}{\Theta_{2}}(s),
\end{eqnarray}
and
\begin{eqnarray}
\mathcal{{F}}_{L}(s,Q^{2})=g(s,Q^{2}){\times}{\Theta_{L}}(s).
\end{eqnarray}
In Eqs. (15) and (16) I used the fact that the Laplace transform
of a convolution function is simply  ordinary product of the
Laplace transform of that function. Taking into account the gluon
distribution, one could extend the Laplace  transformation  to the
high-order corrections in the following form
\begin{eqnarray}
\mathcal{{F}}_{2}(s,Q^{2})=g(s,Q^{2}){\times}[{\Theta^{LO}_{2}}(s)+\frac{\alpha_{s}}{4\pi}{\Theta^{NLO}_{2}}(s)+...],
\end{eqnarray}
and
\begin{eqnarray}
\mathcal{{F}}_{L}(s,Q^{2})=g(s,Q^{2}){\times}[{\Theta^{LO}_{L}}(s)+\frac{\alpha_{s}}{4\pi}{\Theta^{NLO}_{L}}(s)+...].
\end{eqnarray}
The leading-order splitting functions at Laplace $s$- space are
given by:
\begin{eqnarray}
{\Theta^{LO}_{2}}(s)=\frac{1}{1+s}-\frac{2}{2+s}+\frac{2}{3+s},
\end{eqnarray}
and
\begin{eqnarray}
{\Theta^{LO}_{L}}(s)=\frac{1}{2+s}-\frac{1}{3+s}.
\end{eqnarray}
I have, therefore, the derivative of the structure function in the
form of the longitudinal structure function in $s$-space as
\begin{eqnarray}
\frac{\partial F_{2}(s,Q^{2})}{\partial
\ln(Q^{2})}=\frac{1}{4}h(s)F_{L}(s,Q^{2}),
\end{eqnarray}
or
\begin{eqnarray}
F_{L}(s,Q^{2})=4h^{-1}(s)\frac{\partial F_{2}(s,Q^{2})}{\partial
\ln(Q^{2})},
\end{eqnarray}
where $h(s)=\frac{{\Theta^{LO}_{2}}(s)}{{\Theta^{LO}_{L}}(s)}$.\\
The inverse Laplace transforms of  $h(s)$ and $h^{-1}(s)$ is given
by kernels
$\widehat{\eta}(\nu){\equiv}{\mathcal{L}}^{-1}[h(s);\nu]$ and
$\widehat{J}(\nu){\equiv}{\mathcal{L}}^{-1}[h^{-1}(s);\nu]$.
Therefore I have
\begin{eqnarray}
\widehat{\eta}(\nu)=2\delta(\nu)+\delta'(\nu)+2e^{-\nu},
\end{eqnarray}
and
\begin{eqnarray}
\widehat{J}(\nu)=e^{-\frac{3}{2}\nu}\cos(\frac{1}{2}\sqrt{7}\nu)-\frac{1}{7}\sqrt{7}e^{-\frac{3}{2}\nu}\sin(\frac{1}{2}\sqrt{7}\nu).
\end{eqnarray}
Consequently, the general relation between the structure functions
in $x$-space is given by
\begin{eqnarray}
\frac{\partial F_{2}(x,Q^{2})}{\partial
\ln(Q^{2})}&=&\frac{1}{2}F_{L}(x,Q^{2})-\frac{1}{4}x\frac{\partial
F_{L}(x,Q^{2})}{\partial
x}\nonumber\\
&&+\frac{1}{2}\int_{x}^{1}F_{L}(y,Q^{2})\frac{x}{y^{2}}dy.
\end{eqnarray}
Finally, one could write the leading-order relation for the
longitudinal structure function for massless quarks in the form of
the derivative of the structure function as
\begin{eqnarray}
F_{L}(x,Q^{2})|_{N_{f}=4}&=&4\int_{x}^{1}\frac{\partial
F_{2}(y,Q^{2})}{\partial
\ln(Q^{2})}(\frac{x}{y})^{3/2}[\cos(\frac{\sqrt{7}}{2}\ln{\frac{y}{x}})\nonumber\\
&&-\frac{\sqrt{7}}{7}\sin(\frac{\sqrt{7}}{2}\ln{\frac{y}{x}})]\frac{dy}{y}.
\end{eqnarray}

\textit{B:Three flavours + Heavy:} The heavy quark contribution
(charm and bottom) to relation between $F_{L}$ and $F_{2}$ is
define  by fixed-order number scheme using $m_{c}=1.5~GeV$ and
$m_{b}=4.5~GeV$ [8]. The mass of these heavy quarks satisfies
$m_{Q}{\gg}\Lambda _{QCD}$, and  provides a hard scale for pQCD
calculations. One could consider the perturbative predictions for
the longitudinal structure function.\\
 Eq. (2) can be rewritten as
the convolution form
\begin{eqnarray}
x^{-1}F_{L}(x,Q^{2})&{=}&<e^{2}>|_{N_{f}=3}C_{L,g}(\alpha_{s},x)\otimes
g(x,Q^{2})\nonumber\\
&&+x^{-1}F_{L}^{c}+x^{-1}F_{L}^{b},
\end{eqnarray}
where $\otimes$ in the $N_{f}=3$ for massless  quarks u, d and s
denotes the common convolution, and $F_{L}^{c(b)}$ are heavy quark
corrections to the longitudinal structure function at small $x$.\\
These corrections in deep inelastic electron-proton scattering
collisions serve as a test of pQCD and  the heavy quark production
is directly sensitive to the gluon density and  heavy-quarks mass.
One should write the individual longitudinal structure functions
as
\begin{eqnarray}
F_{L}(x,Q^{2})=F_{L}^{g}+F_{L}^{heavy}.
\end{eqnarray}
At small $x$, where the gluon distribution is dominant, the heavy
quark contributions $F_{k}^{i}(x,Q^{2},m^{2}_{i})$ with $i=b,c$
and $k=2,L$  in the proton structure function is written as
\begin{eqnarray}
F_{k}^{i}(x,Q^{2},m^{2}_{i})&=&C_{g,k}^{i}(x,Q^{2}){\otimes}g(x,Q^{2})\nonumber\\
&&=2xe_{i}^{2}\frac{\alpha_{s}(\mu^{2})}{2\pi}\int_{ax}^{1}\frac{dy}{y}C_{g,k}^{i}(\frac{x}{y},Q^{2})g(y,\mu^{2}),\nonumber\\
\end{eqnarray}
where $a=1+4\frac{m_{i}^{2}}{Q^{2}}$, and the renormalization
scale
 $\mu$ is assumed to be average
$<\mu^{2}>=4m_{i}^{2}+\frac{Q^{2}}{2}$.\\
 Using the Laplace
transform method [7], one can rewrite the heavy structure
functions in terms of the convolution integrals with respect to
$\nu'$ and $\omega'$ variables at small $x$ as
\begin{eqnarray}
\hat{FF_{k}^{i}}(\frac{1}{a}\nu',Q^{2})=\int_{0}^{\nu'}\hat{G}(\omega',Q^{2})\\\nonumber
\frac{1}{a}e^{-(\nu'-\omega')}{C_{g,k}^{i}}(\frac{1}{a}e^{-(\nu'-\omega')})dw,
\end{eqnarray}
where
\begin{eqnarray}
\hat{H}_{k}^{i}(\nu')&{\equiv}&
\frac{1}{a}e^{-\nu'}{C_{g,k}^{i}}(\frac{1}{a}e^{-(\nu')}).
\end{eqnarray}
Here $\nu'={\ln}\frac{1}{ax}$, $\omega'={\ln}\frac{1}{ay}$ and
\begin{eqnarray}
\hat{FF_{k}^{i}}(\nu',Q^{2}) &{\equiv}&
(2\frac{\alpha^{LO}_{s}(\mu^{2})}{2\pi}e^{2}_{i})^{-1}\hat{F_{k}^{i}}.
\end{eqnarray}
The Laplace transformation of  $\hat{H_{k}^{i}}(\nu')$ is given by
$h_{k}^{i}(s)$, where
\begin{eqnarray}
h_{k}^{i}(s)&{\equiv}&
{\mathcal{L}}[\hat{H_{k}^{i}}(\nu');s]=\int_{0}^{\infty}\hat{H_{k}}(\nu')e^{-s\nu'}d\nu'.
\end{eqnarray}
The convolution theorem for Laplace transforms allows one to
rewrite the heavy distribution functions as a product of their
Laplace transforms $g(s,Q^{2})$ and $h_{k}^{i}(s)$. In this case
one has
\begin{eqnarray}
FF_{2}^{i}(s,Q^{2})&=&{\mathcal{L}}[\int_{0}^{\nu'}\hat{G}(\omega',Q^{2})\hat{H_{2}^{i}}(\nu'-\omega');s]\\\nonumber
&&=g(s,Q^2)h_{2}^{i}(s),
\end{eqnarray}
and
\begin{eqnarray}
FF_{L}^{i}(s,Q^{2})&=&{\mathcal{L}}[\int_{0}^{\nu'}\hat{G}(\omega',Q^{2})\hat{H_{L}^{i}}(\nu'-\omega');s]\\\nonumber
&&=g(s,Q^2)h_{L}^{i}(s).
\end{eqnarray}
Therefore the ratio of the heavy structure functions are
independent of the gluon distribution function in $s$-space. This
ratio can be written as
\begin{eqnarray}
\frac{F_{L}^{i}(s)}{F_{2}^{i}(s)}=\frac{h_{L}^{i}(s)}{h^{i}_{2}(s)}.
\end{eqnarray}
If one takes the inverse Laplace transformation of Eq. (36), then
one has
\begin{eqnarray}
F_{L}^{i}(\nu')={\mathcal{L}}^{-1}[{F_{2}^{i}(s)}{Rh^{i}(s)};\nu'],
\end{eqnarray}
where $Rh^{i}(s)=\frac{h_{L}^{i}(s)}{h_{2}^{i}(s)}$.\\
 Here I
used the following property for inverse Laplace transformation
\begin{eqnarray}
{\mathcal{L}}^{-1}[F(s)G(s)]&=&\int_{0}^{t}f(t-\tau)g(\tau)d\tau\nonumber\\
&&=\int_{0}^{t}g(t-\tau)f(\tau)d\tau.
\end{eqnarray}
Then Eq. (37) becomes as
\begin{eqnarray}
F_{L}^{i}(\nu',Q^{2})
=\int_{0}^{\nu'}{F_{2}^{i}(\omega',Q^{2})}{\widehat{J}^{i}(\nu'-\omega')}dw',
\end{eqnarray}
where
${\widehat{J}^{i}(\nu')}={\mathcal{L}}^{-1}[{Rh^{i}(s)};\nu']$.\\
The analytical results for the  parameters $\widehat{J}^{i}$ for a
particular range of $Q^{2}$ under study are given in Appendix A.\\
In a similar manner, the longitudinal structure function can be
determined at small $x$  by considering the heavy corrections to
the structure function. Thus, applying the convolution theorem,
the analytical solution for the longitudinal structure function
for $N_{f}=3+\mathrm{Heavy}$ should be converted to usual
$(x,Q^{2})$ space. Therefore one has
\begin{widetext}
 \begin{eqnarray}
F_{L}(x,Q^{2})|_{N_{f}=3+Heavy}=\frac{16}{5}\int_{x}^{1}\frac{\partial
F_{2}(y,Q^{2})}{\partial
\ln(Q^{2})}(\frac{x}{y})^{3/2}[\cos(\frac{\sqrt{7}}{2}\ln{\frac{y}{x}})-\frac{\sqrt{7}}{7}\sin(\frac{\sqrt{7}}{2}\ln{\frac{y}{x}})]\frac{dy}{y}\nonumber\\
+\int_{ax}^{1}\frac{dy}{y}F_{2}^{c}(y,Q^{2})J^{c}(\frac{x}{y},Q^{2})+\int_{ax}^{1}\frac{dy}{y}F_{2}^{b}(y,Q^{2})J^{b}(\frac{x}{y},Q^{2}).
 \end{eqnarray}
 \end{widetext}
one observes that the connection between the  structure functions
(in Eqs. (26) and (40)) are independent of the running coupling
constant at LO analysis and gluon density behavior. To calculate
the right hand side of these equations (Eqs. (26) and (40)) one
has to have an expression for the proton structure function [9]
and heavy quark structure
[10] functions for massless and heavy quarks .\\
The H1 Collaboration reported a measurement of inclusive $ep$
cross sections at high $Q^{2}$ at $\sqrt{s} = 225$ and $252~GeV$.
HERA provided the first measurements of  $F_{L}$ in the region
$120 \leq Q^{2} \leq 800~ GeV^{2}$ and $6.5 {\times} 10^{-4} < x <
0.032$ [1]. My results are compared with  extracted longitudinal
proton structure function $F_{L}$ in the range of  $6.5 \leq Q^{2}
\leq 800~ GeV^{2}$.\\
 In Fig.1 the determined longitudinal
structure function  $F_{L}$ is shown for $ Q^{2}=20~ GeV^{2}$ and
$200~ GeV^{2}$, respectively. In this figure, the longitudinal
structure functions determined for four massless quarks at
$m_{c}^{2}<\mu^{2}$ and also to account for fixed $N_{f}=3$ flavor
number scheme as the heavy flavor contributions to $F_{L}$ are
taken as given by fixed order perturbation theory. The results for
$Q^{2}=20$ and $200~GeV^{2}$ are presented for $N_{f}=4$ and
$N_{f}=3+\mathrm{Heavy}$, and are compared with $H1$ Collaboration
data [1]. For heavy contributions to $F_{L}$, the renormalization
scale is $<\mu^{2}>=4m^{2}_{H}+Q^{2}/2$. These results are
accompanied with errors due to fit parametriztions of $\partial
F_{2}/\partial { \ln}Q^{2}$, as listed in   Appendix B and Table
I. It is seen from Fig.1 that the results are comparable with the
experimental data as accompanied with total errors, although those
are independent of the gluon behavior. I also present the
Cooper-Prytz (CP) fit [3-4], which depends on expanding of the
gluon distribution at $z=1/2$, and the Gay Ducati-Boroun (BG) fit
[5-6] which depends on expanding of the gluon distribution at
$z=\alpha$.\\
 In Fig.2 I present the longitudinal structure
function $F_{L}$ for $Q^{2}=45$ and $500~GeV^{2}$ without
considering  the heavy quark contributions in the same Fig.1. The
longitudinal proton structure function $F_{L}(x, Q^{2})$ compared
by averaging $F_{L}$ data from Table 5 in Ref.[1] at the given
values of $Q^{2}$ and $x$ with total uncertainty on $F_{L}$, shown
in Fig.3. A reasonable agreement between the longitudinal
structure function  as extracted from the direct measurement of
the derivative of $F_{2}$ with the
experimental data is observed at moderate and high $Q^{2}$ values at low values of $x$.\\
These results extend from the LO up to NNLO analysis with respect
to the Laplace transform method at small $x$ and I will try to compare our result with experimental data in the next section.\\

\subsection{III. High-order corrections}
An analytical solution based on the Laplace transformation for the
relation between the longitudinal structure function in terms of a
convolution of the derivative of $F_{2}$ obtained at LO accuracy
in perturbative QCD in section II. Some analytical solutions of
the DGLAP evolution equation in next-to-leading order (NLO)
analysis using the Laplace transform method have been presented in
Ref.[11]. In Refs.[12-13], the authors  have been reported the
complete two and three-order coefficient functions for the
longitudinal structure functions in deep-inelastic scattering
(DIS). Now, a detailed analysis has been performed in order to
find an analytical solutions of the longitudinal structure
function into the derivative of the proton structure function with
respect to ${\ln}Q^{2}$ , using the repeated Laplace
transform, at NLO up to NNLO approximation.\\
In $s$-space, one can rewrite the gluonic structure functions
equations in terms of the convolution integrals up to NNLO
analysis. The Laplace transform of these equations converted to an
ordinary first order differential equations in $s$- space as one
has
\begin{eqnarray}
\frac{\partial{\ln}F_{2}(s,Q^{2})}{\partial{\ln}Q^{2}}=\frac{5}{18}\frac{\alpha_{s}(Q^{2})}{4\pi}\int_{0}^{\nu}(\widehat{P}^{LO}(\nu-\omega)
+\frac{\alpha_{s}(Q^{2})}{4\pi}\nonumber\\
\widehat{P}^{NLO}(\nu-\omega)+(\frac{\alpha_{s}(Q^{2})}{4\pi})
^{2}\widehat{P}^{NNLO}(\nu-\omega))\widehat{G}(\omega,Q^{2})d\omega\nonumber\\
=\frac{5}{18}\frac{\alpha_{s}(Q^{2})}{4\pi}(\Theta_{2}^{LO}(s)
+\frac{\alpha_{s}(Q^{2})}{4\pi}
\Theta_{2}^{NLO}(s)+(\frac{\alpha_{s}(Q^{2})}{4\pi})
^{2}\nonumber\\
\Theta_{2}^{NNLO}(s))g(s,Q^{2}),~~~~~
\end{eqnarray}
and
\begin{eqnarray}
F_{L}(s,Q^{2})=<e^{2}>\frac{\alpha_{s}(Q^{2})}{4\pi}\int_{0}^{\nu}(\widehat{c}^{LO}(\nu-\omega)
+\frac{\alpha_{s}(Q^{2})}{4\pi}\nonumber\\
\widehat{c}^{NLO}(\nu-\omega)+(\frac{\alpha_{s}(Q^{2})}{4\pi})
^{2}\widehat{c}^{NNLO}(\nu-\omega))\widehat{G}(\omega,Q^{2})d\omega\nonumber\\
=<e^{2}>\frac{\alpha_{s}(Q^{2})}{4\pi}(\Theta_{L}^{LO}(s)
+\frac{\alpha_{s}(Q^{2})}{4\pi}
\Theta_{L}^{NLO}(s)+(\frac{\alpha_{s}(Q^{2})}{4\pi})
^{2}\nonumber\\
\Theta_{L}^{NNLO}(s))g(s,Q^{2}).\nonumber\\
\end{eqnarray}
where the running coupling constants have the following forms in
NLO and NNLO analysis respectively as
\begin{eqnarray}
\frac{\alpha_{s}^{\rm
NLO}}{4\pi}=\frac{1}{\beta_{0}t}[1-\frac{\beta_{1}{\ln}t}{\beta_{0}^{2}t}],
\end{eqnarray}
and
\begin{eqnarray}
\frac{\alpha_{s}^{\rm
NNLO}}{4\pi}&=&\frac{1}{\beta_{0}t}[1-\frac{\beta_{1}{\ln}t}{\beta_{0}^{2}t}+\frac{1}{(\beta_{0}t)^{2}}
[(\frac{\beta_{1}}{\beta_{0}})^{2}\nonumber\\
&&(\ln^{2}t-{\ln}t+1)+\frac{\beta_{2}}{\beta_{0}}]].
\end{eqnarray}
where $\beta_{0}=\frac{1}{3}(33-2n_{f})$,
$\beta_{1}=102-\frac{38}{3}n_{f}$ and
$\beta_{2}=\frac{2857}{6}-\frac{6673}{18}n_{f}+\frac{325}{54}n_{f}^{2}$
are the one-loop,two-loop and three-loop corrections to the QCD
$\beta$-function and $\Lambda$ is the QCD cut- off parameter. The
$\Lambda_{QCD}$ parameter usually defined  at NLO and NNLO
analysis as $\Lambda^{(N_{f} =4)}_{QCD}=347~ MeV$ and
$\Lambda^{(N_{f} =4)}_{QCD}=251~ MeV$ [12-13], respectively.\\
The $\mathrm{N^{n}LO}$ expansion coefficients are defined in Ref.
[13] in Mellin-space and one should  present these splitting
functions and  coefficient functions in Appendix C. In detail the
shape of the structure functions are dominated by the gluon
density at low values of $x$. Therefore one would find
\begin{eqnarray}
F_{L}(x,Q^{2})={\mathcal{L}}^{-1}[H(s,Q^{2})DF_{2}(s,Q^{2});\nu],
\end{eqnarray}
where
$DF_{2}{\equiv}\frac{{\partial}F_{2}}{{\partial}{\ln}Q^{2}}$. The
high-order $H(\nu,Q^{2})$ for four $Q^{2}$ values presented in
Appendix D. One can easily determine these high order corrections
to the gluonic longitudinal structure function based on the
derivative of the proton structure function with respect to $\ln
Q^{2}$ at low $x$. Now considering the terms from NLO up to NNLO,
the gluonic longitudinal structure function takes the following
form for a given $Q^{2}$ value as
\begin{widetext}
\begin{eqnarray}
F_{L}^{NLO}(x,Q^2)|_{Q^{2}=20~GeV^{2}}&=&-0.07DF_{2}(x,Q^2)+\int_{x}^{1}\frac{dy}{y}DF_2(y,Q^{2})
(-0.13(\frac{x}{y})^{0.08}+(\frac{x}{y})^{1.54}
[3.73\cos(1.33\ln(\frac{x}{y}))\nonumber\\
&&-1.68(\sin(1.33\ln(\frac{x}{y}))]),\nonumber\\
F_{L}^{NNLO}(x,Q^2)|_{Q^{2}=20~GeV^{2}}&=&-0.31DF_{2}(x,Q^2)+\int_{x}^{1}\frac{dy}{y}DF_2(y,Q^{2})
((\frac{y}{x})^{0.06}[-0.3\cos(0.26\ln(\frac{x}{y}))+0.11\sin(0.26\ln(\frac{x}{y}))]\nonumber\\
&&+(\frac{x}{y})^{1.40}[3.82\cos(1.31\ln(\frac{x}{y}))-1.17(\sin(1.31\ln(\frac{x}{y}))]).
\end{eqnarray}
\end{widetext}
For my numerical investigation, the high order corrections to
$F_{L}(x,Q^{2})$ are shown in Fig.4 and compared with H1 data [1]
for $Q^{2}=20,~45,~200~ \mathrm{and}~ 500~ GeV^{2}$. In this
figure the straight and dash lines represent the gluonic
longitudinal structure function solutions at NLO and NNLO
respectively. These results are obtained with respect to  the
Laplace transform technique as described in Appendix D. In this
figure, the circles represent the longitudinal structure functions
from Ref. [1] as accompanied with total errors. These results are
in agrement with $F_{L}(x,Q^{2})$ predicted from the global fit at
LO, NLO and NNLO in Ref. [12]. However it is a reflection of the
behavior of the deep inelastic structure function and the
coefficient functions at low values of $x$.\\
In the next sections, the recombination processes between gluons
in a dense system have to be taken into account. Therefore the
gluonic longitudinal structure function behavior has to be tamed
by
screening effects.\\

\subsection{IV. Nonlinear behavior}

The screening effects are provided by a multiple gluon interaction
which leads to the nonlinear terms in the derivation of the linear
DGLAP evolution equations. Therefore the standard linear DGLAP
evolution equations will have to be  modified in order to take the
nonlinear corrections into account.\\
 Gribov, Levin, Ryskin,
Mueller and Qiu (GLR-MQ) [14] performed a detailed study of these
recombination processes. This  widely known as the GLR-MQ equation
and involves the two-gluon distribution per unit area of the
hadron. This equation predicts a saturation behavior of the gluon
distribution at very small $x$ [15-16]. A closer examination of
the small $x$ scattering is resummation powers of
$\alpha_{s}\ln(1/x)$ where leads to the $k_{T}$-factorization form
[17]. In the $k_{T}$-factorization approach the large logarithms
$\ln(1/x)$ are relevant for the unintegrated gluon density in a
nonlinear equation.  Solution of this equation develops a
saturation scale where tame the gluon density
behavior at low values of $x$ and this is an intrinsic characteristic of a dense gluon system.\\
Therefore one should consider the low- $x$ behavior of the singlet
distribution using the nonlinear GLR-MQ evolution equation. The
shadowing correction to the evolution of the singlet quark
distribution can be written as [13-14,18]
\begin{eqnarray}
\frac{{\partial}xq(x,Q^{2})}{{\partial}{\ln}Q^{2}}=\frac{{\partial}xq(x,Q^{2})}{{\partial}{\ln}Q^{2}}|_{DGLAP}-\frac{27\alpha_{s}^{2}}{160R^{2}Q^{2}}
[xg(x,Q^{2})]^{2}.\nonumber\\
\end{eqnarray}
Eq. (47) can be rewrite in a convenient form as
\begin{eqnarray}
\frac{{\partial}F_{2}(x,Q^{2})}{{\partial}{\ln}Q^{2}}=\frac{{\partial}F_{2}(x,Q^{2})}{{\partial}lnQ^{2}}|_{DGLAP}-
\frac{5}{18}\frac{27\alpha_{s}^{2}}{160R^{2}Q^{2}}\nonumber\\
{\times}[xg(x,Q^{2})]^{2}.
\end{eqnarray}
The first term is the standard DGLAP evolution equation (Eq. 11)
and the value of $R$ is the correlation radius between two
interacting gluons. It will be  of the order of the proton radius
$(R\simeq5\hspace{0.1cm} GeV^{-1})$, if the gluons are distributed
through the whole of proton, or much smaller
$(R\simeq2\hspace{0.1cm} GeV^{-1})$ if gluons are concentrated in
hot- spot within the proton.\\
One would find Eq. (48) at LO analysis in $s$-space as
\begin{eqnarray}
\frac{{\partial}F_{2}(s,Q^{2})}{{\partial}{\ln}Q^{2}}=\frac{10\alpha_{s}}{18\pi}G(s,Q^{2})\Theta^{LO}_{2}(s)-
\frac{5}{18}\frac{27\alpha_{s}^{2}}{160R^{2}Q^{2}}\nonumber\\
{\times}G^{2}(s,Q^{2}).
\end{eqnarray}
The longitudinal structure function in $s$-space is given in the
following form
\begin{eqnarray}
F_{L}(s,Q^{2})=\frac{20\alpha_{s}}{9\pi}G(s,Q^{2})\Theta^{LO}_{L}(s).
\end{eqnarray}
Combining Eqs. (49) and (50), one could calculate the nonlinear
relation between the derivative of the structure function and
longitudinal structure function in $s$-space as I have
\begin{eqnarray}
\frac{{\partial}F_{2}(s,Q^{2})}{{\partial}{\ln}Q^{2}}=\frac{h(s)}{4}F_{L}(s,Q^{2})-\frac{\zeta}{\Theta^{2}_{L}(s)}F_{L}^{2}(s,Q^{2}),
\end{eqnarray}
where $\zeta=\frac{243\pi^2}{25600R^{2}Q^{2}}$. At $\zeta
{\rightarrow}~ 0$, Eq. (51) leads to the linear relation between the structure functions (i.e., Eq. 21).\\
Eq. (51) yields the gluonic longitudinal structure function with
nonlinear effects as
\begin{eqnarray}
F_{L}^{2}(s,Q^{2})-\frac{h(s)}{4}\frac{\Theta^{2}_{L}(s)}{\zeta}F_{L}(s,Q^{2})+\frac{\Theta^{2}_{L}(s)}{\zeta}DF_{2}(s,Q^{2})\nonumber\\
=0.~~~~
\end{eqnarray}
It is tempting, however, one of the roots of Eq. (52) can be
discarded. Solution of Eq.(52) then leads us to a solution for the
nonlinear gluonic longitudinal structure function. This equation
can be solved by Taylor series expansion method around a
particular choice of point of expansion. Since $(\zeta
R^{2}Q^{2})^{n}{<}1$, so this series is convergent when
$n{\rightarrow}\infty$. This parameter decreases with increasing
$n$, as seen from Table II. For the longitudinal structure
function in $s$-space, one has
\begin{eqnarray}
F_{L}(s,Q^{2})&=&4h^{-1}(s)DF_{2}(s,Q^{2})+64{\zeta}\frac{\Theta_{L}(s)}{\Theta^{3}_{2}(s)}DF_{2}^{2}(s,Q^{2})\nonumber\\
&&+2048{\zeta^{2}}\frac{\Theta_{L}(s)}{\Theta^{5}_{2}(s)}DF_{2}^{3}(s,Q^{2})...~.
\end{eqnarray}
Eq. (53) covers the whole range of expanding as it is shown in
Table II. The contribution from the fourth term to the second term
(such that
$\frac{\mathrm{Fourth~term}}{\mathrm{Second~term}}{\propto}\frac{\zeta^{3}}{\zeta}=\zeta^{2}$)
is around the order of $\mathcal{O}(\sim10^{-2})$. To make a rough
estimate of the accuracy in expansion method I find the
longitudinal structure function until four order approximation
with respect to the $\zeta$ expansion and neglecting the high
order terms $\mathcal{O}(\zeta^{3})$ in Eq. (53). For this
evolution, I retain the second order term into $\zeta$. Therefore
the gluonic longitudinal structure function in $\nu$-space is
defined as
\begin{eqnarray}
\widehat{F}_{L}(\nu,Q^{2})&=&4\int_{0}^{1}\widehat{DF}_{2}(\nu,Q^{2})\widehat{J}(\nu-\omega)d\omega\\\nonumber
&&+\int_{0}^{1}\widehat{DF}^{2}_{2}(\nu,Q^{2})\widehat{P}(\nu-\omega)d\omega\nonumber\\
&&+\int_{0}^{1}\widehat{DF}^{3}_{2}(\nu,Q^{2})\widehat{T}(\nu-\omega)d\omega,
\end{eqnarray}
where
$\widehat{J}(\nu){\equiv}{\mathcal{L}}^{-1}[h^{-1}(s);\nu]{\equiv}\frac{1}{4}W_{1}^{LO}(\nu)$,
$\widehat{P}(\nu){\equiv}{\mathcal{L}}^{-1}[64{\zeta}\frac{\Theta_{L}(s)}{\Theta^{3}_{2}(s)};\nu]{\equiv}W_{2}^{LO}(\nu,Q^{2})$
and
$\widehat{T}(\nu){\equiv}{\mathcal{L}}^{-1}[2048{\zeta^{2}}\frac{\Theta_{L}(s)}{\Theta^{5}_{2}(s)};\nu]{\equiv}W_{3}^{LO}(\nu,Q^{2})$.\\
The inverse-Laplace transform of kernels can be found in Appendix
E. Applying the properties of Dirac delta function, finally I have
the nonlinear gluonic longitudinal structure function in $x$-space
by the following form
 \begin{widetext}
 \begin{eqnarray}
F_{L}(x,Q^{2})|_{N_{f}=4}=\mathrm{Eq.(26)}+\int_{x}^{1}\widehat{DF}^{2}_{2}(\nu,Q^{2})W_{2}^{LO}(\nu-\omega,Q^{2})d\omega
+\int_{0}^{1}\widehat{DF}^{3}_{2}(\nu,Q^{2})W_{3}^{LO}(\nu-\omega,Q^{2})d\omega,
\end{eqnarray}
\end{widetext}
  where $W_{1}^{LO}$ is independent of the values of $Q^{2}$, but
 $W_{2}^{LO}$ and $W_{3}^{LO}$ are depend on $Q^{2}$ values.
Thus I obtained an expression for the gluonic longitudinal
structure function $F_{L}(x,Q^{^{2}})$ in leading order by solving
the nonlinear GLR-MQ evolution equation. Eq. (56) shows that it is
independent of the gluon behavior, the running coupling constant,
and also the QCD cut off parameter in the LO approximation. One
can easily solve this equation (i.e., Eq. 56), and extract the
nonlinear gluonic longitudinal structure function.\\
 The nonlinear
behavior of  $F_{L}(x,Q^{2})$ is shown in Fig.5 for values of
$Q^{2}=6.5$ and $20~GeV^{2}$. It would appear that the effect of
nonlinearity at low-$x$ values should observe for moderate $Q^{2}$
values when compared with H1 data. In this figure, the nonlinear
effect investigated at hot-spot point ($R=2~ GeV^{-1}$). It is
shown that the obtained results from present analysis based on
Laplace transform are in good agreements with the ones obtained by
H1 Collaboration [1]. The saturation of the gluon density at small
$x$ indirectly is significant for understanding the nonlinear
effects in Eq. (56) and also high-order corrections. In the next section I apply high order corrections to the nonlinear behavior and compared with H1 data.\\

\subsection{V. High-order corrections to the nonlinear behavior}
Using the formalism given in the previous section, I  calculate
the high-order corrections to the nonlinear behavior of the
gluonic longitudinal structure function at low $x$ region. In
terms of the derivative of proton structure function with respect
to $\ln Q^{2}$, the GLR-MQ evaluation equation can be written in
the high order correction in $s$-space as
\begin{eqnarray}
\frac{{\partial}F_{2}(s,Q^{2})}{{\partial}{\ln}Q^{2}}=\frac{5}{18}\frac{\alpha_{s}(Q^{2})}{4\pi}\Theta_{2}(s,Q^{2})g(s,Q^{2})\nonumber\\
-\frac{5}{18}\frac{27\alpha_{s}^{2}(Q^{2})}{160R^{2}Q^{2}}g^{2}(s,Q^{2}),
\end{eqnarray}
where $\Theta_{2}(s,Q^{2})=\Theta_{2}^{LO}(s)
+\frac{\alpha_{s}(Q^{2})}{4\pi}
\Theta_{2}^{NLO}(s)+(\frac{\alpha_{s}(Q^{2})}{4\pi})^{2}\Theta_{2}^{NNLO}(s)$.
One should consider the same method introduce in the previous
section, I find the high order corrections to the gluonic
longitudinal structure function by the following form
\begin{eqnarray}
\frac{{\partial}F_{2}(s,Q^{2})}{{\partial}{\ln}Q^{2}}=\frac{5}{18}\frac{\Theta_{2}(s,Q^{2})}{<e^{2}>\Theta_{L}(s,Q^{2})}F_{L}(s,Q^{2})\nonumber\\
-\frac{3\pi^{2}}{4R^{2}Q^{2}}\frac{F_{L}^{2}(s,Q^{2})}{(<e^{2}>\Theta_{L}(s,Q^{2}))^{2}},
\end{eqnarray}
where $\Theta_{L}(s,Q^{2})=\Theta_{L}^{LO}(s)
+\frac{\alpha_{s}(Q^{2})}{4\pi}
\Theta_{L}^{NLO}(s)+(\frac{\alpha_{s}(Q^{2})}{4\pi})^{2}\Theta_{L}^{NNLO}(s)$.
Eq.(58) can be solved simultaneously to get  the desired nonlinear
equation for longitudinal structure function. Using the inverse
Laplace transform to back from $s$-space to $x$-space, the
simplified solution of the above equation at high-order
corrections can be obtained by
\begin{eqnarray}
F_{L}(x,Q^{2})=Eq.(45)+{\mathcal{L}}^{-1}[\frac{B(s,Q^{2})}{A^{3}(s,Q^{2})}D^{2}F_{2}(s,Q^{2});\nu]\nonumber\\
+{\mathcal{L}}^{-1}[2\frac{B^{2}(s,Q^{2})}{A^{5}(s,Q^{2})}D^{3}F_{2}(s,Q^{2});\nu],~~~
\end{eqnarray}
where
$A(s,Q^{2})=\frac{5}{18}\frac{\Theta_{2}(s,Q^{2})}{<e^{2}>\Theta_{L}(s,Q^{2})}$
and
$B(s,Q^{2})=\frac{3\pi^{2}}{4R^{2}Q^{2}(<e^{2}>\Theta_{L}(s,Q^{2}))^{2}}$.\\
Therefore the solution of the nonlinear corrections at NLO up to
NNLO analysis leads us to nonlinear behavior of the gluonic
longitudinal structure function at moderate values of $Q^{2}$. The
analytical expressions for these corrections are given in Appendix
F. The validity of the nonlinear corrections to the DGLAP
evolution equation is in the region of small $x$ and intermediate
values of $Q^{2}$. The nonlinear corrections can be neglected at
large values of $Q^{2}$, so I expect that my result to be valid in
the kinematic region $x{\leq}0.01$ and moderate
$Q^{2}$.\\
In Fig.6 the nonlinear behavior for moderate and high $Q^{2}$
values are shown. One would expect that this behavior to be
observe at moderate $Q^{2}$ values as consider in Fig.6.  From
these figures, it is observed that the NLO nonlinear corrections
(NLO+NLCs) show tamed behavior to those obtained from only NLO
corrections when compared with H1 data. Its observed that NNLO
nonlinear corrections (NNLO+NLCs) has negative rate as $x$
decreases at moderate
$Q^{2}$ values.\\
 Indeed, comparison of the NNLO+NLCs with the
NNLO calculations shows a turnover of the gluonic longitudinal
structure function at $Q^{2}=6.5$ and $20~GeV^{2}$. This is due to
the effect of the gluonic coefficient function to the gluonic
splitting function ratio , which decreases the limit NNLO
corrections when tamed with respect to the nonlinear saturation
effect. Since gluon recombination introduces a negative correction
to the NNLO linear behavior, the signal of its presence is a
decrease of the scaling violation and this is strongly dependence
to the correlation radius (i.e., R). In Fig.7,  the effect of the
nonlinearity in NNLO results for $R=4~GeV^{-1}$ at $Q^{2}=6.5$ and
$20~GeV^{2}$ investigated. It can be observed (in Figs.6 and 7)
that NNLO results are very sensitive to $R$ as $x$ decreases.
Indeed the effect third-order corrections to the coefficients
functions and splitting functions in hot-spots point decrease the
gluonic longitudinal structure function to the negative values as
$x$ decreases. This behavior is comparable when
$R$ increase throughout the entire proton at NNLO approximation.\\

At least there is another mechanism to prevent generation of the
high density gluon states, as this is well known the vacuum color
screening [19]. There is a transition between the nonperturbative
and perturbative domains. In the QCD vacuum, the non-perturbative
fields form structures with sizes $\sim R_{c} $ which it is
smaller than $\Lambda_{QCD}$. The short propagation
length for perturbative gluons is $R_{c}\sim 0.2-0.3~fm$.\\
The gluon fusion effect in non-linear regime controlled by the new
dimensionless parameter $\sim \frac{R_{c}^{2}}{8B}$ where $B$ is
the characteristic size of the interaction region as this
parameter can be defined by $\ln(x_{0}/x)$ and $r$ where $r^{2}
\sim Q^{-2}$. In all figures one should observe that the nonlinear
effects are small even at lowest $x$ values. This behavior is in
accordance with the smallness of the ratio $
\frac{R_{c}^{2}}{8B}$. It is interesting to look at the nonlinear
limit where decreases as $Q^{2}$ increases. From [19], the
nonlinear effects leads to the logarithmically ratio as the
nonlinear/linear effects are proportional to
$R_{c}^{2}/8B(\ln(x_{0}/x),r^{2})\ln(Q^{2}R^{2}_{c})$. Figs.6 are
shown that high order corrections to the nonlinear behavior are
very small at high $Q^{2}$ values and at lowest available Bjorken
$x$.
\subsection{VI. Conclusion}
In this paper I have estimated an analytical solution for
 the linear and nonlinear behaviors of the longitudinal structure
 function with respect to the derivative of the proton structure
 function inside the proton.\\
  This solution is independent of the
 gluon model and the running coupling constant at leading order
 analysis and it is  free of any point expanding  model for the gluon
 distribution behavior. The ratio of splitting functions applying the Laplace transform technique are calculated.
  I have used the heavy coefficient functions for
 heavy-flavour production in DIS in the fixed-flavour-number scheme (FFNS) with
 $N_{f}=3$. In the  present calculations the high-order corrections (NLO and NNLO)
 for structure functions at low $x$ values,
 arising from the coefficient functions and the splitting
 functions, are obtained. I have therefore used from these results for the gluonic longitudinal
 structure function at moderated and high values of $Q^{2}$.\\
  The nonlinear GLR-MQ evolution equation
 predicted by considering the general Laplace transform method and studied
 the effects of adding the nonlinear corrections to the linear
 longitudinal structure function at hot-spot point ($R=2~GeV^{-1}$) with $N_{f}=4$.
 For the gluonic longitudinal structure function the nonlinear effects are
 found to play an increasingly important role at $x\leq 10^{-3}$. I have incorporated high-order corrections to the nonlinear
 behavior in the kinematic range of moderate-$Q^{2}$ and obtained
 the nonlinear longitudinal structure function at low $x$ at NLO
 and NNLO approximation. It is interesting to see that the NNLO
 analysis at moderate $Q^{2}$ is dependence to the proton radius as the nonlinear behavior increase as $R$
 increases. This is due to the contribution from the NNLO terms in
 the ratio of coefficient function to the splitting function. It
 can be observed that with decreasing $x$, the taming of
 $F_{L}(x,Q^{2})$ is apparently observed in NLO approximation at
 $R=2~GeV^{-1}$ and in NNLO approximation at $R=4~GeV^{-1}$. This method presented in this analysis enable us to
 achieve strictly analytical linear and nonlinear solutions at LO up to NNLO approximation
  for the gluonic longitudinal structure function in terms of the derivative of the proton structure function with respect to the
  ${\ln}Q^{2}$ at low values of $x$. The nonlinear effects are shown
to be small at large $Q^{2}$, even at
lowest Bjorken values of $x$.\\

\subsection{Acknowledgment}
Author thanks Urs A.Wiedemann for discussions which completed this
study and the Department of Physics of the CERN-TH for their warm
hospitality. Also I would like to thank M.Tabrizi for his
careful editorial revising of the paper.\\
\subsection{ Appendix. A}
The kernels at leading order analysis are as follows:
\begin{eqnarray}
P_{qg}^{LO}(z)=z^2+(1-z)^2,
 \end{eqnarray}
  and
  \begin{eqnarray}
c^{LO}_{L,g}(z)=z^{2}(1-z).
 \end{eqnarray}
 The parameters $\widehat{J}^{c}(\upsilon)$ and $\widehat{J}^{b}(\upsilon)$ are given
 by the following form:
 At $Q^{2}=20~GeV^{2}$
 \begin{widetext}
 \begin{eqnarray}
\widehat{J}^{c}(\upsilon)=.252\exp(-3.183v)\cos(.198v)+.424\exp(-3.183v)\sin(.198v)-.0232\exp(-2.050v)\nonumber\\
-.0624\exp(-1.127v)\cos(.305v)-.0692\exp(-1.127v)\sin(.305v)
+.156\delta(v),\nonumber\\
\widehat{J}^{b}(\upsilon)=-4.035\exp(-7.471v)-.0745\exp(-2.246v)+.0445\exp(-1.313v)\cos(.393v)\nonumber\\
-.154\exp(-1.313v)\sin(.393v)+.370\exp(-.310v)+1.405\delta(v).
 \end{eqnarray}
 \end{widetext}
 At $Q^{2}=200~GeV^{2}$
 \begin{widetext}
 \begin{eqnarray}
\widehat{J}^{c}(\upsilon)=-.0164\exp(-3.169v)+.111exp(-2.609v)-.003\exp(-2.005v)\nonumber\\
-.0912\exp(-1.164v)\cos(.218v)-.0293\exp(-1.164v)\sin(.218v)+.228\delta(v),\nonumber\\
\widehat{J}^{b}(\upsilon)=.222\exp(-3.146v)\cos(.152v)+.518\exp(-3.146v)\sin(.152v)-.0212\exp(-2.046v)\nonumber\\
-.0645\exp(-1.127v)\cos(.299v)-.0685\exp(-1.127v)\sin(.299v)+.171\delta(v).
 \end{eqnarray}
 \end{widetext}
\subsection{ Appendix. B}
The proton structure function parameterized with a  global fit
function [9] to the HERA combined data for $F^{\gamma p}_{ 2}
(x,Q^{2})$ for $0.85< Q^{2} < 3000~ GeV^{2}$ and $x< 0.1$, which
ensures that the saturated Froissart $\ln^{2}(1/x)$ behavior
dominates at small-$x$. This global fit takes the form
\begin{eqnarray}
F^{\gamma p}_{ 2} (x,Q^{2})& =& (1- x)[\frac{F_{P}}{1 -x_{P}} +
A(Q^{2})\ln( \frac{x_{P}}{ x}\frac{1 - x}{ 1 - x_{P}})\nonumber\\
&&+B(Q^{2}) \ln^{2}( \frac{x_{P}}{ x}\frac{1 - x}{ 1 - x_{P}})],
\end{eqnarray}
where
\begin{eqnarray}
 A(Q^{2}) = a_{0} + a_{1} {\ln}Q^{2} + a_{2} {\ln}^{2}
 Q^{2},\nonumber
 \end{eqnarray}
and
\begin{eqnarray}
  B(Q^{2}) = b_{0} + b_{1}
{\ln}Q^{2} + b_{2} {\ln}^{2} Q^{2}.\nonumber
\end{eqnarray}
The fitted parameters are tabulated in Table I. At small $x$ ( or
large $\nu = \ln(1/x)$), the global fit becomes a quadratic
polynomial in $\nu$ as

$ \widehat{F}^{\gamma p}_{ 2} (\nu,Q^{2}){\rightarrow} C_{0f}
(Q^{2}) +  C_{1f} (Q^{2})\nu + C_{2f} (Q^{2})\nu^{2} +
\widehat{O}(\nu)$ where the coefficient functions are defined in
Ref. [9].\\
 \subsection{ Appendix. C}
At small $x$ the one-loop up to three-loop splitting functions for
$N_{f}=4$ read
\begin{eqnarray}
P^{LO}=2N_{f}(1-2x+2x^{2}),\nonumber\\
P^{NLO}{\rightarrow}C_{A}T_{f}\frac{40}{9x},~~~~~~~~~~~`\nonumber\\
P^{NNLO}{\rightarrow}E_{1}^{qg}\frac{\ln
x}{x}+E_{2}^{qg}\frac{1}{x},~~
\end{eqnarray}
 where $E_{1}^{qg}{\simeq}-298.667N_{f}$ and
 $E_{2}^{qg}{\simeq}-1268.28N_{f}+4.57613N_{f}^{2}$.
The gluonic longitudinal coefficient functions up to NNLO analysis
at small $x$ can be written as
\begin{eqnarray}
c^{LO}=8N_{f}x(1-x),~~~~~~~~~~~~~~~~~~~~~~~~~~~~~~~~~~~~~~~\nonumber\\
c^{NLO}{\rightarrow}\frac{-5.333N_{f}}{x}+(-6.229N_{f}+0.8889N_{f}^{2})~~~~~~~,\nonumber\\
c^{NNLO}{\rightarrow}N_{f}(\frac{-2044.70}{x}-409.506\frac{\ln x
}{x})+N_{f}^{2}\frac{88.5037}{x}.\nonumber\\
\end{eqnarray}
 \subsection{ Appendix. D}
The high-order ratios for some of $Q^{2}$ values at NLO and NNLO
analysis are
\begin{widetext}
\begin{eqnarray}
H^{NLO}(\nu,20)&=&\exp(-1.54\nu)(3.73\cos(1.33\nu)-1.68\sin(1.33v))-
0.13\exp(-0.82E-1\nu) -0.68E-1\delta(\nu),\nonumber\\
H^{NNLO}(\nu,20)&=&\exp(-1.40\nu)(3.82\cos(1.31\nu)-1.17\sin(1.31v))
+\exp(+0.59\nu)(-0.30\cos(.26\nu)+0.11\sin(.26v))\nonumber\\
&&-0.31\delta(\nu).\nonumber\\
H^{NLO}(\nu,45)&=&\exp(-1.54\nu)(3.77\cos(1.33\nu)-1.66\sin(1.33v))
- 0.12\exp(-0.73E-1\nu)-0.60E-1\delta(\nu),\nonumber\\
H^{NNLO}(\nu,45)&=&\exp(-1.43\nu)(3.84\cos(1.32\nu)-1.25\sin(1.32v))
+\exp(+0.04\nu)(-0.24\cos(.23\nu)+0.01\sin(.23v))\nonumber\\
&&-0.25\delta(\nu).\nonumber\\
H^{NLO}(\nu,200)&=&\exp(-1.53\nu)(3.81\cos(1.33\nu)-1.64\sin(1.33v))
- 0.10\exp(-0.61E-1\nu)-0.50E-1\delta(\nu),\nonumber\\
H^{NNLO}(\nu,200)&=&\exp(-1.45\nu)(3.87\cos(1.32\nu)-1.35\sin(1.32v))
+\exp(+0.22E-1\nu)(-0.17\cos(.19\nu)+0.85\sin(.19v))\nonumber\\
&&-0.18\delta(\nu).\nonumber\\
H^{NLO}(\nu,500)&=&\exp(-1.53\nu)(3.83\cos(1.33\nu)-1.63\sin(1.33v))
- 0.91E-1\exp(-0.55E-1\nu)-0.45E-1\delta(\nu),\nonumber\\
H^{NNLO}(\nu,500)&=&\exp(-1.46\nu)(3.88\cos(1.32\nu)-1.39\sin(1.32v))
+\exp(+0.16E-1\nu)(-0.15\cos(.17\nu)\nonumber\\
&&+0.78E-1\sin(.17v))-0.15\delta(\nu).
\end{eqnarray}
\end{widetext}

 \subsection{ Appendix. E}
The inverse-Laplace of the nonlinear kernels are as follows:
\begin{eqnarray}
\widehat{J}(\nu)=Eq.(26),\nonumber
 \end{eqnarray}
 \begin{widetext}
\begin{eqnarray}
\widehat{P}(\nu)&=&\zeta((-56768/343\sqrt{7}+3328/49\sqrt{7}\nu+1408/49\nu^2\sqrt{7})\sin(1/2\sqrt{7}\nu)\exp(-3/2\nu)\nonumber\\
&&+(-20224/49\nu+128/7\nu^2-320)\exp(-3/2\nu)\cos(1/2\sqrt{7}\nu)+64\delta'(\nu)+256\delta(\nu)),\nonumber\\
\widehat{T}(\nu)&=&\zeta^2((145276928/2401\sqrt{7}\nu-40165376/2401\nu^2\sqrt{7}+203331584/2401\sqrt{7}
-3866624/1029\nu^3\sqrt{7}\nonumber\\
&&+20480/1029\nu^4\sqrt{7})\sin(1/2\sqrt{7}\nu)\exp(-3/2\nu)
+(-3604480/1029\nu^3+63062016/343\nu+22896640/343\nu^2\nonumber\\
&&-69632/147\nu^4-100352)\exp(-3/2\nu)\cos(1/2\sqrt{7}\nu)
+2048\delta(3,\nu)+20480\delta(2,\nu)+49152\delta(1,\nu)-61440\delta(\nu)).\nonumber\\
W_{1}^{LO}(\nu,6.5)&=&-0.57\exp(-1.5\nu)(-7\cos(1.32\nu)+2.64\sin(1.32\nu)),\nonumber\\
W_{2}^{LO}(\nu,6.5)&=&0.92\delta(\nu)+0.23\delta(1,\nu)+\exp(-1.5\nu)((-1.153-1.49\nu+0.066\nu^{2})\cos(1.32\nu)\nonumber\\
&&+(-1.58+0.65\nu+0.27\nu^{2})\sin(1.32\nu)),\nonumber\\
W_{3}^{LO}(\nu,6.5)&=&-0.8\delta(\nu)+0.64\delta(1,\nu)+0.27\delta(2,\nu)+0.026\delta(3,\nu)\nonumber\\
&&+\exp(-1.5\nu)((-1.3+2.39\nu+0.87\nu^{2}-0.045\nu^{3}-0.006\nu^{4})\cos(1.32\nu)\nonumber\\
&&+(2.91+2.08\nu-0.57\nu^{2}-0.13\nu^{3}+0.0007\nu^{4})\sin(1.32\nu)),\nonumber\\
W_{1}^{LO}(\nu,20)&=&\mathrm{Constant}(W_{1}^{LO}(\nu,6.5)),\nonumber\\
W_{2}^{LO}(\nu,20)&=&0.30\delta(\nu)+0.075\delta(1,\nu)+\exp(-1.5\nu)((-0.37-0.48\nu+0.02\nu^{2})\cos(1.32\nu)\nonumber\\
&&+(-0.51+0.21\nu+0.09\nu^{2})\sin(1.32\nu)),\nonumber\\
W_{3}^{LO}(\nu,20)&=&-0.084\delta(\nu)+0.067\delta(1,\nu)+0.03\delta(2,\nu)+0.003\delta(3,\nu)\nonumber\\
&&+\exp(-1.5\nu)((-0.14+0.25\nu+0.09\nu^{2}-0.005\nu^{3}-0.0006\nu^{4})\cos(1.32\nu)\nonumber\\
&&+(0.31+0.22\nu-0.061\nu^{2}-0.014\nu^{3}+0.00007\nu^{4})\sin(1.32\nu)),\nonumber\\
 \end{eqnarray}
 \end{widetext}

 \subsection{ Appendix. F}
 The nonlinear corrections for some of $Q^{2}$ values at NLO up to NNLO
analysis are
\begin{widetext}
\begin{eqnarray}
W_{1}^{NLO}(\nu,6.5)&=&-0.08\delta(\nu)-0.15\exp(-0.1\nu)+\exp(-1.55\nu)(3.67\cos(1.33\nu)-1.70\sin(1.33\nu)),\nonumber\\
W_{2}^{NLO}(\nu,6.5)&=&0.57\delta(\nu)+0.15\delta(1,\nu)-0.004\delta(2,\nu)+\exp(-1.55\nu)((-1.17-1.02\nu+0.13\nu^{2})\cos(1.33\nu)\nonumber\\
&&+(-0.99+0.86\nu+0.2\nu^{2})\sin(1.33\nu))+\exp(-0.1\nu)(-0.016+0.003\nu-0.00006\nu^{2}),\nonumber\\
W_{3}^{NLO}(\nu,6.5)&=&-0.74\delta(\nu)+0.25\delta(1,\nu)+0.14\delta(2,\nu)+0.013\delta(3,\nu)-0.0004\delta(4,\nu)\nonumber\\
&&+\exp(-1.55\nu)((-0.017+2.15\nu+0.35\nu^{2}-0.08\nu^{3}-0.004\nu^{4})\cos(1.33\nu)\nonumber\\
&&+(2.15+0.55\nu-0.67\nu^{2}-0.06\nu^{3}+0.003\nu^{4})\sin(1.33\nu)),\nonumber\\
&&+\exp(-0.1\nu)(-0.0002+0.0006\nu-0.00006\nu^{2}+0.13E-5\nu^{3}-0.84E-8\nu^{4}),\nonumber\\
W_{1}^{NNLO}(\nu,6.5)&=&-0.45\delta(\nu)+\exp(0.1\nu)(-0.50\cos(0.31\nu)+0.10\sin(0.31\nu))+\exp(-1.35\nu)(3.76\cos(1.3\nu)\nonumber\\
&&-0.99\sin(1.3\nu)),\nonumber\\
W_{2}^{NNLO}(\nu,6.5)&=&0.98\delta(\nu)+0.014\delta(1,\nu)-0.036\delta(2,\nu)+\exp(-1.35\nu)((0.12-1.85\nu-0.18\nu^{2})\cos(1.3\nu)\nonumber\\
&&+(-2.38-0.65\nu+0.28\nu^{2})\sin(1.3\nu))+\exp(0.1\nu)((-0.18-0.014\nu+0.002\nu^{2})\cos(0.31\nu)\nonumber\\
&&+(-0.045+0.04\nu+0.001\nu^{2})\sin(0.31\nu)),\nonumber\\
W_{3}^{NNLO}(\nu,6.5)&=&0.76\delta(\nu)+1.2\delta(1,\nu)+0.11\delta(2,\nu)-0.04\delta(3,\nu)-0.006\delta(4,\nu)\nonumber\\
&&+\exp(-1.35\nu)((-5.69-2.3\nu+1.22\nu^{2}+0.2\nu^{3}-0.001\nu^{4})\cos(1.3\nu)\nonumber\\
&&+(-0.22+5\nu+1.23\nu^{2}-0.1\nu^{3}-0.01\nu^{4})\sin(1.3\nu)),\nonumber\\
&&+\exp(0.1\nu)((-0.26-0.011\nu+0.007\nu^{2}+0.0002\nu^{3}-0.12E-5\nu^{4})\cos(0.31\nu)\nonumber\\
&&+(0.047+0.08\nu+0.004\nu^{2}-0.0002\nu^{3}-0.35E-5\nu^{4})\sin(0.31\nu)),
\end{eqnarray}
\end{widetext}
and
\begin{widetext}
\begin{eqnarray}
W_{1}^{NLO}(\nu,20)&=&-0.07\delta(\nu)-0.13\exp(-0.082\nu)+\exp(-1.54\nu)(3.73\cos(1.33\nu)-1.68\sin(1.33\nu)),\nonumber\\
W_{2}^{NLO}(\nu,20)&=&0.20\delta(\nu)+0.054\delta(1,\nu)-0.001\delta(2,\nu)+\exp(-1.54\nu)((-0.38-0.36\nu+0.04\nu^{2})\cos(1.33\nu)\nonumber\\
&&+(-0.35+0.27\nu+0.07\nu^{2})\sin(1.33\nu))+\exp(-0.082\nu)(-0.005+0.0007\nu-0.00001\nu^{2}),\nonumber\\
W_{3}^{NLO}(\nu,20)&=&-0.082\delta(\nu)+0.032\delta(1,\nu)+0.02\delta(2,\nu)+0.002\delta(3,\nu)-0.00004\delta(4,\nu)\nonumber\\
&&+\exp(-1.54\nu)((-0.02+0.24\nu+0.05\nu^{2}-0.008\nu^{3}-0.0004\nu^{4})\cos(1.33\nu)\nonumber\\
&&+(+0.24+0.083\nu-0.07\nu^{2}-0.008\nu^{3}+0.0003\nu^{4})\sin(1.33\nu)),\nonumber\\
&&+\exp(-0.082\nu)(-0.0001+0.00007\nu-0.45E-5\nu^{2}+0.8E-7\nu^{3}-0.4E-9\nu^{4}),\nonumber\\
W_{1}^{NNLO}(\nu,20)&=&-0.31\delta(\nu)+\exp(0.06\nu)(-0.30\cos(0.26\nu)+0.11\sin(0.26\nu))+\exp(-1.4\nu)(3.82\cos(1.31\nu)\nonumber\\
&&-1.17\sin(1.31\nu)),\nonumber\\
W_{2}^{NNLO}(\nu,20)&=&0.32\delta(\nu)+0.034\delta(1,\nu)-0.007\delta(2,\nu)+\exp(-1.4\nu)((-0.15-0.6\nu-0.028\nu^{2})\cos(1.31\nu)\nonumber\\
&&+(-0.71-0.035\nu+0.1\nu^{2})\sin(1.31\nu))+\exp(0.06\nu)((-0.26-0.0006\nu+0.0002\nu^{2})\cos(0.26\nu)\nonumber\\
&&+(-0.01+0.005\nu+0.00001\nu^{2})\sin(0.26\nu)),\nonumber\\
W_{3}^{NNLO}(\nu,20)&=&-0.011\delta(\nu)+0.11\delta(1,\nu)+0.023\delta(2,\nu)-0.0007\delta(3,\nu)-0.0003\delta(4,\nu)\nonumber\\
&&+\exp(-1.4\nu)((-0.45+0.025\nu+0.15\nu^{2}+0.11\nu^{3}-0.0006\nu^{4})\cos(1.31\nu)\nonumber\\
&&+(0.20+0.48\nu+0.042\nu^{2}-0.017\nu^{3}-0.0007\nu^{4})\sin(1.31\nu)),\nonumber\\
&&+\exp(0.06\nu)((-0.008-0.00004\nu+0.0001\nu^{2}+0.13E-4\nu^{3}-0.44E-7\nu^{4})\cos(0.26\nu)\nonumber\\
&&+(0.001+0.002\nu+0.00002\nu^{2}-0.45E-5\nu^{3}-0.23E-7\nu^{4})\sin(0.26\nu)),
\end{eqnarray}
\end{widetext}
where $W^{,}s$ are inverse Laplace transform of all coefficients
at LO up to NNLO analysis accordance with results expanded in nonlinear behavior.\\

 \begin{table}[h]
\caption{ Parameters of Eq. (65), resulting from a global fit to
the HERA combined data.}
\begin{tabular} {cccc}
\toprule \\  \multicolumn{2}{c}{parameters \quad \quad \quad ~~~~~~~~~~~~~~~~value}    \\ &&&\\ \hline \\ &&&\\
  $a_{0} $  &   \quad  $-8.471\times 10^{-2}\pm2.62\times10^{-3} $  \\
  $a_{1} $  &   \quad   $4.190\times 10^{-2}\quad\pm1.56\times10^{-3}$  \\
  $a_{2}$   &    \quad  $-3.976\times 10^{-3}\pm2.13\times10^{-4}$   \\  &&&\\
 $b_{0}$   &   \quad   $1.292\times 10^{-2}\pm3.62\times10^{-4} $ \\

 $b_{1}$   &   \quad   $2.473\times 10^{-4}\pm2.46\times10^{-4}$  \\

 $b_{2}$    &    \quad  $1.642\times 10^{-3}\pm5.52\times10^{-5} $ \\ &&& \\
$F_{p}$& \quad  $0.413\pm 0.003$ & &\\

$\chi^{2}(\mathrm{goodness~ of~ fit})$ &  \quad  $1.17$ & &\\
\hline

\end{tabular}
\end{table}
  \begin{table}[h]
\caption{ Higher order terms in expansion method.}
\begin{tabular}{cccc}
 \toprule \\  \multicolumn{2}{c}{n ~~~~$(\zeta
R^{2}Q^{2})^{n}$}
\\\hline\\
 1  &   \quad   $\mathcal{O}(10^{-1})$  \\
 2  &   \quad   $\mathcal{O}(10^{-2})$  \\
 3  &    \quad  $\mathcal{O}(10^{-3})$   \\
 4  &   \quad   $\mathcal{O}(10^{-4})$ \\
\hline

\end{tabular}
\end{table}
\subsection{References}

1. V. Andreev  et al. [H1 Collab.], Eur. Phys. J. C{\bf74}(2014)2814.\\
2. H. Abramowicz et al. [H1 and ZEUS Collab.], Eur. Phys. J. C {\bf75}(2015)580.\\
3. A. M. Cooper-Sarkar et al., Z. Phys. C{\bf39}(1988)281.\\
4. K. Prytz, Phys. Lett. B{\bf311}(1993)286.\\
5. M. B. Gay Ducati and P. B. Goncalves,
Phys. Lett. B{\bf390}(1997)401.\\
6. G. R. Boroun and B. Rezaei, Eur. Phys. J. C{\bf72}(2012)2221.\\
7. Martin M. Block et al., Phys. Rev. D{\bf79}(2008)014031.\\
8. O. Zenaiev, arXiv:1612.02371 (2016).\\
9. Martin M. Block et al., Phys. Rev. D{\bf88}(2013)014006.\\
10. N. N. Nikolaev and V. R. Zoller, Phys. Lett.
B{\bf509}(2001)283; N. N. Nikolaev and V. R. Zoller, Phys. Atom.
Nucl.{\bf73}(2010)672;
 M. Gluk, C. Pisano and E. Reya, Phys. Rev. D{\bf77}(2008)074002\\
11.M. M. Block et al., Eur. Phys. J. C{\bf69}(2010)425; H.
Khanpour et
al., Phys. Rev. C{\bf95}(2017)035201; S.Shoeibi et al., arXiv:1710.06329 (2017).\\
12. S. Moch et al., Phys. Lett. B{\bf606}(2005)123;
 A. D. Martin et al., Phys. Lett. B{\bf635}(2006)305; Phys. Lett. B{\bf636}(2006)259; R. S. Thorne, arXiv:hep/0511351 (2005).\\
13. A. Vogt et al., Nucl. Phys. B{\bf691}(2004)129; C. D. White and  R. S. Thorne, Eur. Phys. J. C{\bf45}(2006)179.\\
14. A. H. Mueller and J. Qiu, Nucl. Phys. B\textbf{268}(1986)427;
L. V. Gribov, E. M. Levin and M. G. Ryskin, Phys.
Rep.\textbf{100},
 (1983)1.\\
15. G. R. Boroun and B. Rezaei, Chin. Phys. Lett.{\bf32}, (2015)
no.11, 111101; B. Rezaei and G. R. Boroun, Phys. Lett. B{\bf692}
(2010) 247;
 G. R. Boroun, Eur. Phys. J. A{\bf43} (2010) 335.\\
16. M. Devee and J. K. Sarma, Nucl.Phys.B{\bf885}(2014)571;
Eur. Phys. J. C{\bf74}(2014)2751.\\
17. N. N. Nikolaev and W. Sch$\ddot{a}$fer, Phys. Rev. D{\bf74}(2006)014023.\\
18. K. J. Eskola et al., Nucl. Phys. B{\bf660}(2003)211.\\
19. R. Fiore, P. V. Sasorov and V. R. Zoller, JETP Letters
{\bf96}(2013)687; R. Fiore, N. N. Nikolaev and V. R. Zoller,
JETP Letters {\bf99}(2014)363.\\
\begin{figure}
\centering
\includegraphics[width=1\textwidth]{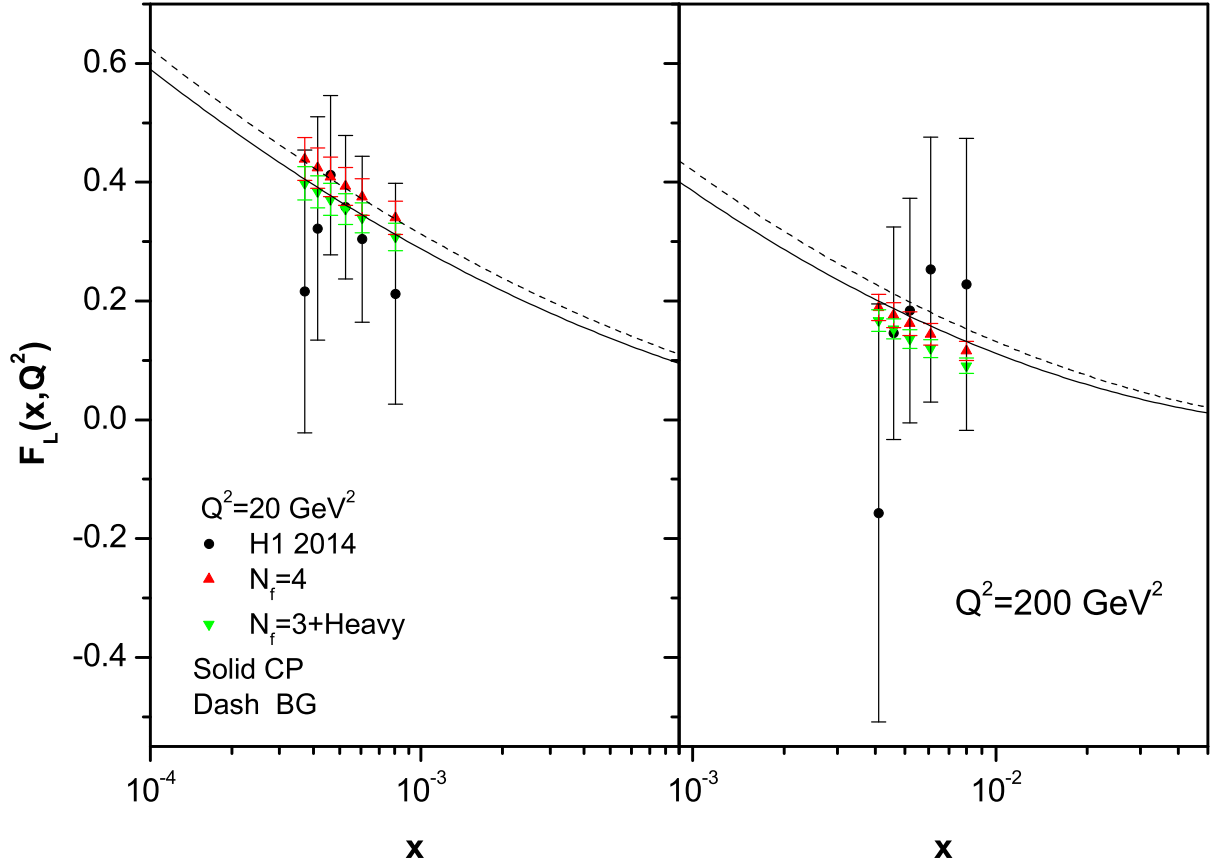}
\caption{The longitudinal structure functions $F_{L} (x, Q^{2})$
(Up triangle ($N_{f}=4$)), Down triangle
($N_{f}=3+\mathrm{Heav}y$)) compared by H1 [1](circles) at the
given values of $Q^{2}$ accompanied with total uncertainties. The
determined error bars represent the derivative of $F_{2}(x,Q^{2})$
uncertainties. The curves represent the prediction from the
expanding of gluon behavior [3-6].} \label{Fig1}
\end{figure}
\begin{figure}
\centering
\includegraphics[width=1\textwidth]{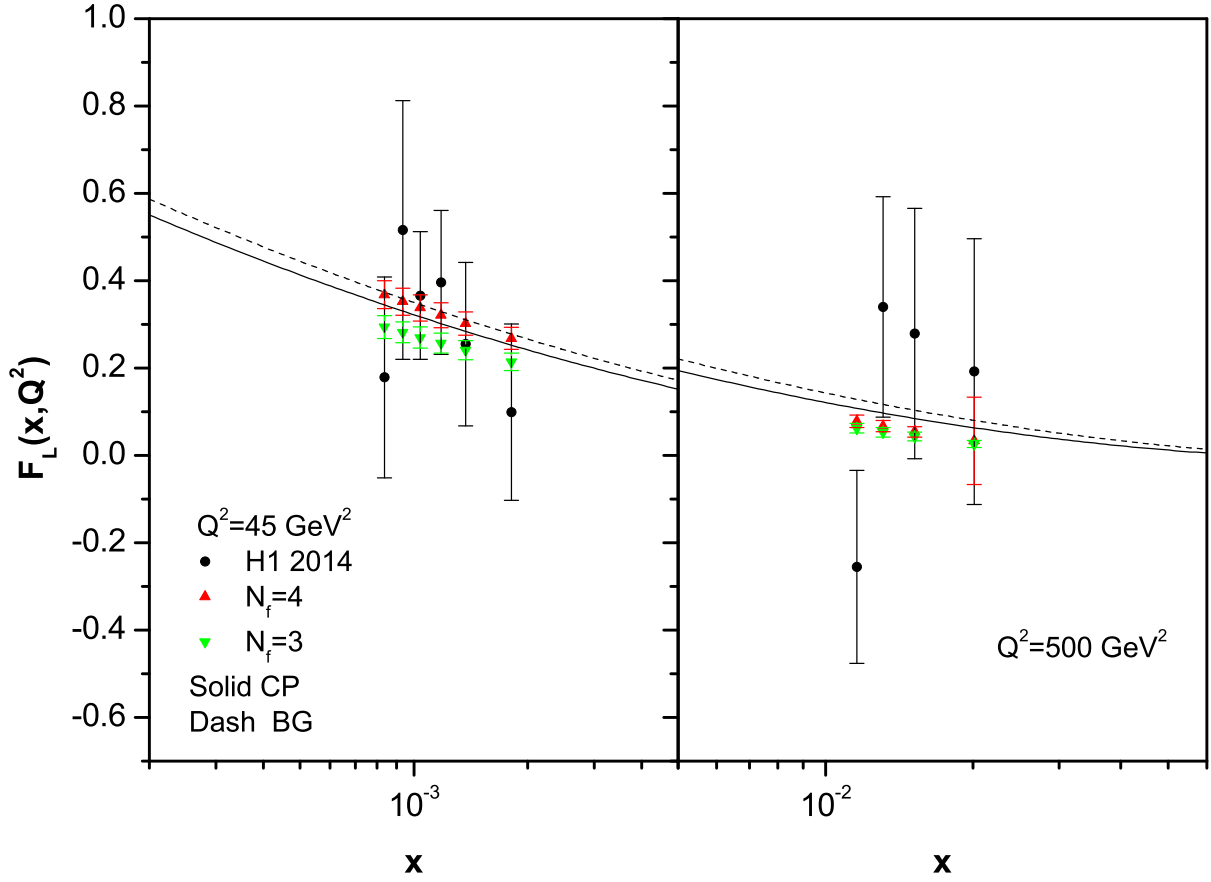}
\caption{ The same as Fig.1 at the given values of $Q^{2}$ for
$N_{f}=4$ and $N_{f}=3$.} \label{Fig2}
\end{figure}
\begin{figure}
\centering
\includegraphics[width=1\textwidth]{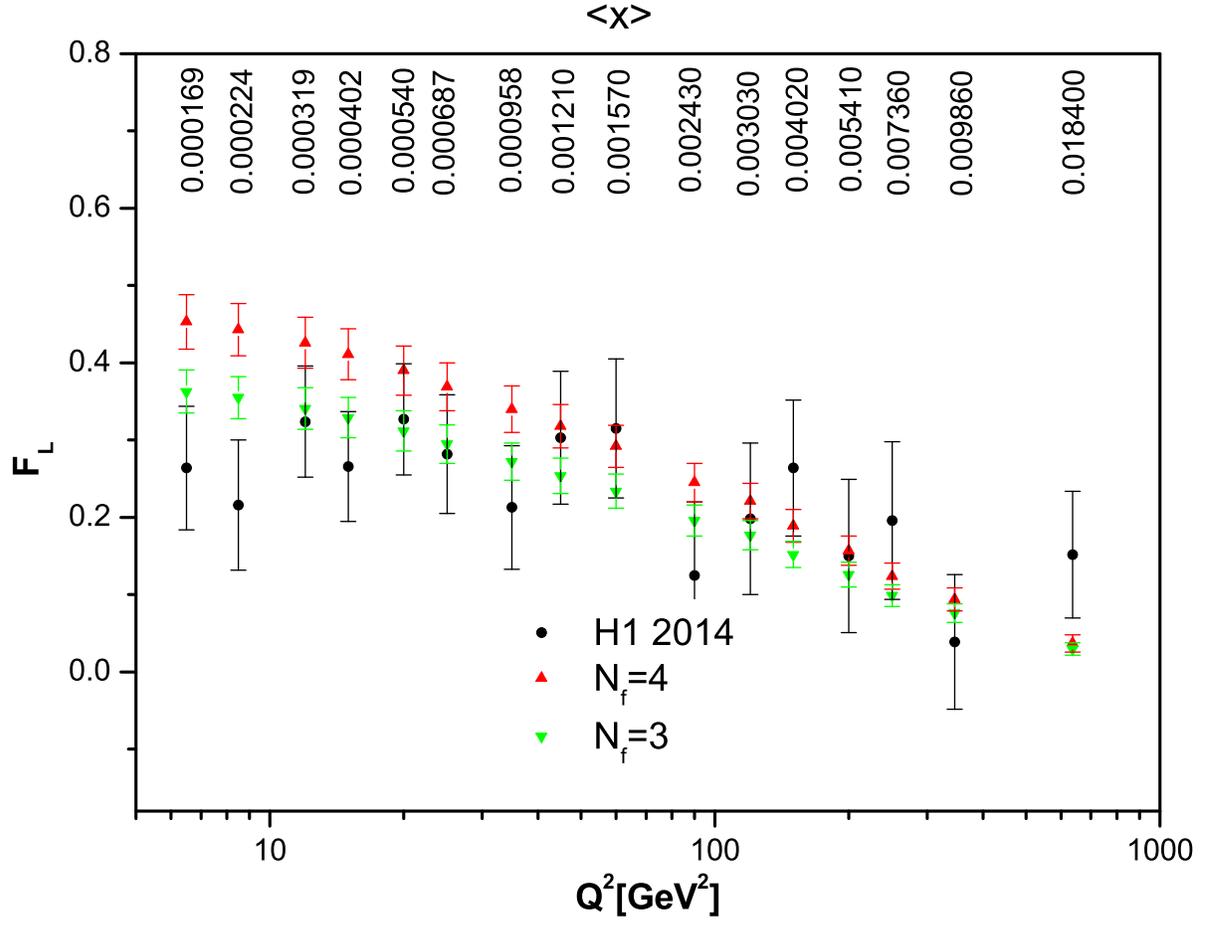}
\caption{The longitudinal structure function $F_{L}$ compared with
H1 data averaged over $x$ in the region $6.5 \leq Q^{2} \leq 800
GeV^{2}$ (solid points). The error bars represent the full errors
as obtained by the Monte Carlo procedure described in the Ref.[1].
For each $Q^{2}$ the average value of $x$ is given  above each
data point. } \label{Fig3}
\end{figure}

\begin{figure}
\centering
\includegraphics[width=0.9\textwidth]{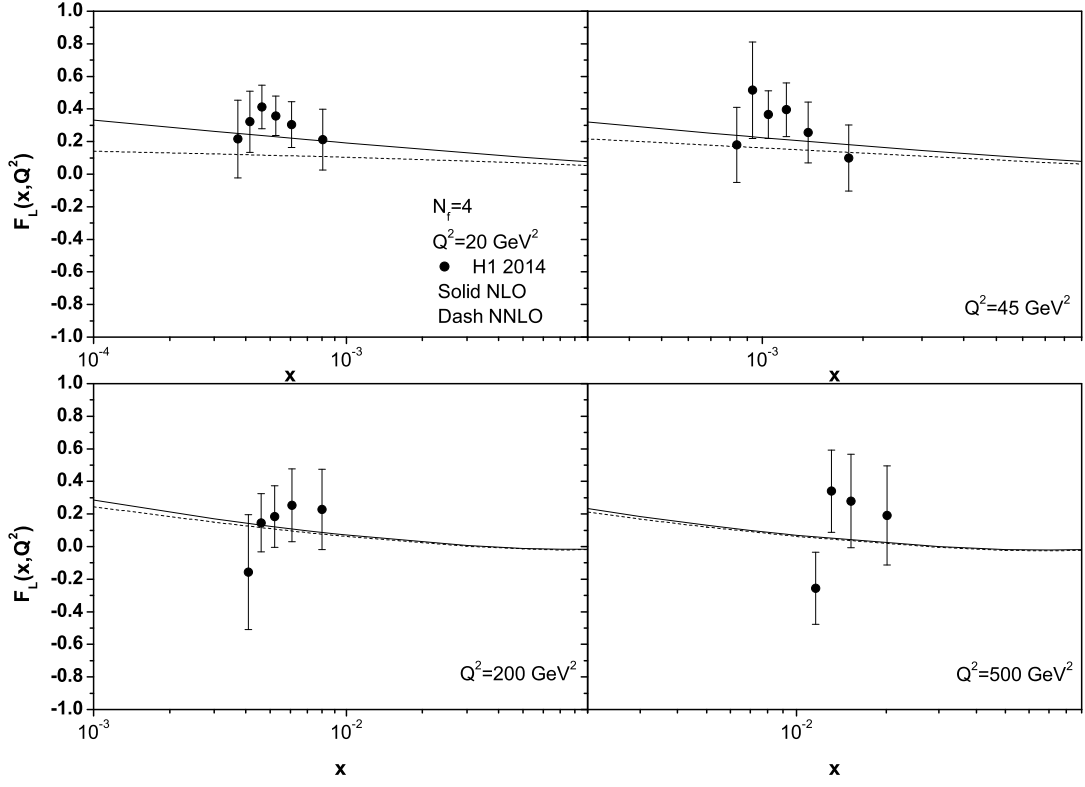}
\caption{The high-order corrections to the gluonic longitudinal
structure function compared with H1 data [1].} \label{Fig4}
\end{figure}

\begin{figure}
\centering
\includegraphics[width=1\textwidth]{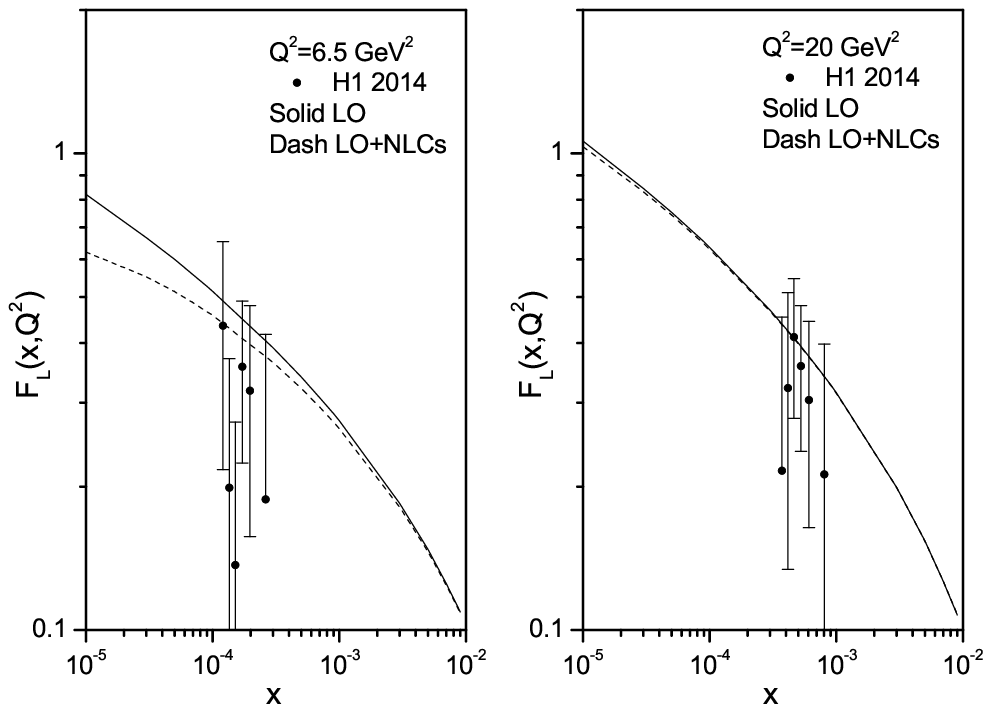}
\caption{Nonlinear corrections (NLCs) to the gluonic longitudinal
structure function $F_{L}$ at LO analysis for $N_{f}=4$ at
$R=2~\mathrm{GeV}^{-1}$ compared with H1 data at $Q^{2}=6.5~
\mathrm{and}~ 20~\mathrm{GeV}^{2}$ (solid points).} \label{Fig5}
\end{figure}
\begin{figure}
\centering
\includegraphics[width=0.9\textwidth]{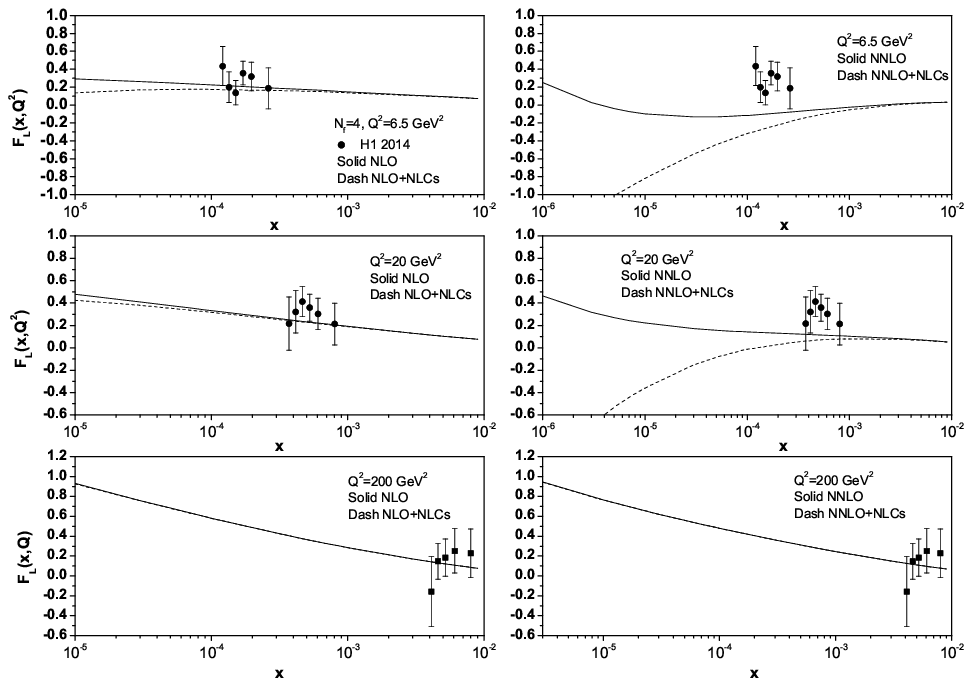}
\caption{High order nonlinear corrections to the gluonic
longitudinal structure function $F_{L}$ at NLO up to NNLO analysis
for $N_{f}=4$ at $R=2~\mathrm{GeV}^{-1}$ compared with H1 data at
$Q^{2}=6.5,~20~ \mathrm{and}~ 200~\mathrm{GeV}^{2}$ (solid
points).} \label{Fig6}
\end{figure}
\begin{figure}
\centering
\includegraphics[width=1\textwidth]{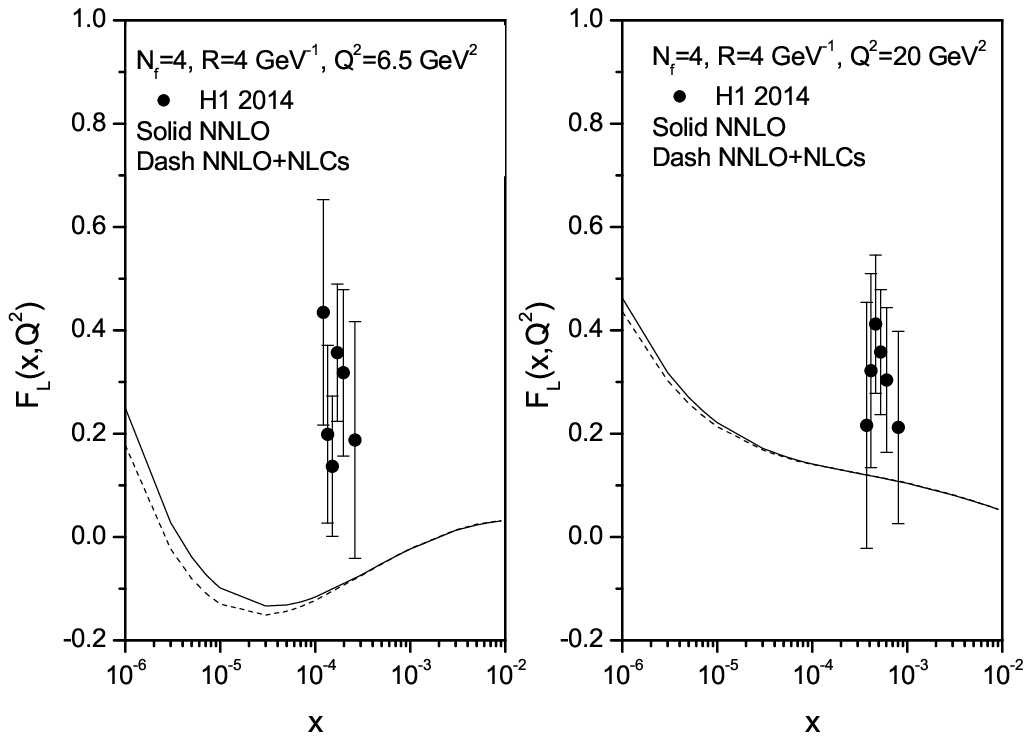}
\caption{NNLO nonlinear corrections to the gluonic longitudinal
structure function $F_{L}$ for $N_{f}=4$ at
$R=4~\mathrm{GeV}^{-1}$ compared with H1 data at $Q^{2}=6.5~
\mathrm{and}~ 20~\mathrm{GeV}^{2}$ (solid points).} \label{Fig7}
\end{figure}

\end{document}